\documentclass[aps,preprint,amsmath,amsfonts,amssymb,showpacs,showkeys]{revtex4}

\usepackage{bm}
\usepackage{graphicx}
\usepackage{epsfig}

\usepackage[colorlinks]{hyperref}

\usepackage[title]{appendix}

\begin{document}

\title{Sum rules for strong decays of 10 and 27 dibaryon multiplets in broken SU(3) symmetry}

\author{G.~S\'anchez-Col\'on}

\email[]{gabriel.sanchez@cinvestav.mx}

\affiliation{ Departamento de F\'{\i}sica Aplicada.\\ Centro de
Investigaci\'on y de Estudios Avanzados del IPN.\\ Unidad
M\'erida.\\ A.P. 73, Cordemex.\\ M\'erida, Yucat\'an, 97310.
MEXICO. }

\author{E.~N.~Polanco-Eu\'an}

\affiliation{ Departamento de F\'{\i}sica Aplicada.\\ Centro de
Investigaci\'on y de Estudios Avanzados del IPN.\\ Unidad
M\'erida.\\ A.P. 73, Cordemex.\\ M\'erida, Yucat\'an, 97310.
MEXICO. }

\author{C.~E.~L\'opez-Fort\'in}

\affiliation{ Facultad de Ingenier\'ia.\\ Universidad Aut\'onoma de Yucat\'an.\\ A.P. 150, Cordemex.\\ M\'erida, Yucat\'an.
MEXICO. }

\date{\today}

\begin{abstract}

\textbf{10} and \textbf{27} SU(3) multiplets of hexaquarks states with baryon number $ B=2 $, known as dibaryons, are considered. Previous analyses are complemented and sum rules for strong decay amplitudes with SU(3) symmetry violation up to the first order are determined for the \textbf{10} and \textbf{27} dibaryon multiplets into a baryon octet plus a baryon decuplet and for the dibaryon \textbf{27} into two baryon octets.
\end{abstract}



\maketitle

\section{Introduction}\label{sec.introduction}

A dibaryon denotes a sextaquark state with baryon number $B=2$. The first known dibaryon has been the deuteron discovered in 1932~\cite{urey32}. In 1963, Oakes~\cite{oakes63} examined the role of deuteron in the \lq\lq eightfold way"~\cite{gellmann61,okubo62} and found that it belongs to the irreducible representation $\bm{10^*}$ resulting from the SU(3) decomposition of two baryon octets, $\bm{8}\otimes\bm{8}$. 

Interest in the study of dibaryons rose in 1977 when Jaffe~\cite{jaffe77} predicted the so-called H-dibaryon, a bound $\Lambda \Lambda$ system containing two strange quarks. Diverse amount of predicted states triggered a worldwide rush of experimental dibaryon searches~\cite{yokosawa80,locheer86,strakovsky91}. Recently, two groups announced that lattice QCD calculations~\cite{beane11,beane13,inoue11} provide evidence for a bound H-dibaryon, as predicted. Despite the historical difficulties in the experimental search for multi-quarks states, tetraquarks ($q_1\bar{q}_2q_3\bar{q}_4$) and pentaquarks ($q_1q_2q_3q_4\bar{q}_5$) systems have been recently observed~\cite{choi03,aaij15}. Moreover, WASA at COSY collaboration have finally confirmed solid evidence for the existence of a dibaryon~\cite{bashkanov,huang}, denoted $d^*(2380)$. Theoretical models and analysis of experimental data have also suggested possible existence of dibaryon states belonging to \textbf{8}~\cite{oka88} and \textbf{27}~\cite{xie84,xie89} SU(3) representations.

It is in this context that the study of sum rules for decay amplitudes turns out to be helpful in the hunting for possible multi-quark states.

A dibaryon has a total of six quarks, for it to be a color singlet it must be a baryon-baryon state and its SU(3) properties depend on the SU(3) properties of the two baryon it is made of. If both baryons are octets, then the SU(3) transformation properties of the dibaryon can be $\bm{1}$, $\bm{8}$, $\bm{10}$, $\bm{10^*}$, or $\bm{27}$. If one of the baryons is an octet and the other one is a decuplet then the dibaryon can transform as $\bm{8}$, $\bm{10}$, $\bm{27}$, or $\bm{35}$.

Flavor SU(3) symmetry breaking effects up to the first order are taken into account in Refs.~\cite{polanco16a,polanco16b} and sum rules are calculated for strong decays of dibaryon octet $\bm{8}$ (denoted $D_{\bm{8}}$) into two baryon octets (denoted $B_{\bm{8}}$), $D_{\bm{8}}\to B_{\bm{8}}+B_{\bm{8}}$, and into a baryon octet plus a baryon decuplet (baryon resonances, denoted $R_{\bm{10}}$), $D_{\bm{8}}\to B_{\bm{8}}+R_{\bm{10}}$. Corresponding sum rules for decays of dibaryon SU(3) decuplets, $\bm{10^*}$ and $\bm{10}$ (denoted $D_{\bm{10^*,10}}$, respectively), into two baryon octets, $D_{\bm{10^*,10}}\to B_{\bm{8}}+B_{\bm{8}}$, are reported in Ref.~\cite{gupta15}.

In this paper, analyses of Refs.~\cite{polanco16a,polanco16b,gupta15} are complemented. Sum rules for strong decays amplitudes with first order breaking of the SU(3) symmetry are determined for dibaryon multiplets \textbf{10} ($D_{\bm{10}}$) and \textbf{27} (denoted $D_{\bm{27}}$) into a baryon octet plus a baryon decuplet, $D_{\bm{10,27}}\to B_{\bm{8}}+R_{\bm{10}}$.

Additionally, sum rules in broken SU(3) symmetry up to first order are determined for strong decays of $D_{\bm{27}}$ into two baryon octets, $D_{\bm{27}}\to B_{\bm{8}}+B_{\bm{8}}$.

Using the traditional method, the SU(3) symmetry breaking interaction term transforms as the eight component of the unitary spin, that is, like the hypercharge~\cite{gellmann61,okubo62}.

In Sec.~\ref{sec.gralformula}, to establish notation and for practical use, the general formula for the two body decay amplitude including SU(3) first order breaking is determined. Secs.~\ref{sec.d10br} and \ref{sec.d27br} include the analysis of $D_{\bm{10,27}}\to B_{\bm{8}}+R_{\bm{10}}$ decays, respectively. $D_{\bm{27}}\to B_{\bm{8}}+B_{\bm{8}}$ decays are studied in Sec.~\ref{sec.d27bb}. Finally, a summary of results obtained and concluding remarks are presented in Sec.~\ref{sec.conclusions}.

\section{Decay amplitude in broken SU(3) general formula}\label{sec.gralformula}

Consider a two body decay, $a\to b+c$, of an initial state $|a\rangle\equiv|\bm{\mu}_a,n_a\rangle$ into $|b\rangle\equiv|\bm{\mu}_b,n_b\rangle$ plus $|c\rangle\equiv|\bm{\mu}_c,n_c\rangle$, where $\bm{\mu}$ is the belonging representation of the corresponding state and $n=(Y,I,I_3)$ collectively denotes its quantum numbers of hypercharge, isospin, and the third component of isospin, respectively. The decay amplitude is given by the matrix element

\begin{eqnarray}
G\left[a\to b\, c\right] \equiv (\langle c|\langle b|)H_{\rm int} |a\rangle & = &
(\langle\bm{\mu}_c,n_c|\langle\bm{\mu}_b,n_b|)H_{\rm int} |\bm{\mu}_a,n_a\rangle
\nonumber\\
& = &
\sum_{\bm{\mu}_{\gamma},n}
\left(
\begin{array}{cc}
\bm{\mu}_b & \bm{\mu}_c \\
n_b & n_c
\end{array}
\right|
\left.
\begin{array}{c}
\bm{\mu}_{\gamma} \\ n
\end{array}
\right)
\langle\bm{\mu}_{\gamma},n| H_{\rm int} |\bm{\mu}_a,n_a\rangle,
\label{ampdecgral}
\end{eqnarray}

\noindent
where $H_{\rm int}$ is the interaction Hamiltonian. Final composite state has been expanded by using SU(3) Clebsch-Gordan coefficients following notation of Ref.~\cite{mcnamee64}. Representation $\bm{\mu}_{\gamma}$ goes through all representations appearing in the $\bm{\mu}_b\otimes \bm{\mu}_c$ decomposition and quantum numbers $n$ are obtained from $n_b$ and $n_c$ by $Y=Y_b+Y_c$, $I=|I_b-I_c|, \ldots,I_b+I_c$, and $I_3=I_{3\,b}+I_{3\,c}$.

In the flavor SU(3) symmetry limit, that is, with $H_{\rm int}= H_{\rm st}$, the matrix element appearing in second line of general expression~(\ref{ampdecgral}) satisfies

\begin{equation}\label{ampsimlim}
\langle\bm{\mu}_{\gamma},n| H_{\rm st} |\bm{\mu}_a,n_a\rangle =
g^{0}_{\bm{\mu}_{\gamma}}\,
\delta_{\bm{\mu}_{\gamma},\bm{\mu}_a}\,
\delta_{n,n_a},
\end{equation}

\noindent
where $g^{0}_{\bm{\mu}_{\gamma}}$ is a coupling constant parameter.

To apply the octet \lq\lq spurion" formalism in this case, define a fictitious state as the $Y=0$, $I=0$ component of the SU(3) octet, $|S_{\bm{8}}\rangle \equiv |\bm{8},n_S\rangle$, with $n_S=(0,0,0)$. Then, up to first order, SU(3) symmetry breaking effects are absorbed into the interaction of initial state with $|S_{\bm{8}}\rangle $ and the original decay, $a\stackrel{H_{\rm ms}}{\longrightarrow}b + c$, can then be treated as an SU(3) invariant \lq\lq dispersion", $S_{\bm{8}} + a\stackrel{H_{\rm st}}{\longrightarrow} b + c$.

With $H_{\rm int}= H_{\rm ms}$, the matrix element in second line of (\ref{ampdecgral}) is then given by:

\begin{eqnarray}
\langle\bm{\mu}_{\gamma},n| H_{\rm ms} |\bm{\mu}_a,n_a\rangle & = & \langle\bm{\mu}_{\gamma},n| H_{\rm st} (|S_{\bm{8}}\rangle|\bm{\mu}_a,n_a\rangle)
\nonumber\\
& = &
\sum_{\bm{\nu}_{\tau},n'}
\left(
\begin{array}{cc}
\bm{8} & \bm{\mu}_a \\
n_S & n_a
\end{array}
\right|
\left.
\begin{array}{c}
\bm{\nu}_{\tau} \\ n'
\end{array}
\right)
\langle\bm{\mu}_{\gamma},n| H_{\rm st}
|\bm{\nu}_{\tau},n'\rangle
\nonumber\\
& = &
{\displaystyle\sum_{\bm{\nu}_{\tau}}}
\left(
\begin{array}{cc}
\bm{8} & \bm{\mu}_a \\
n_S & n_a
\end{array}
\right|
\left.
\begin{array}{c}
\bm{\nu}_{\tau} \\ n_a
\end{array}
\right)
g_{\bm{\nu}_{\tau}}\,
\delta_{\bm{\mu}_{\gamma},\bm{\nu}_{\tau}}\,
\delta_{n,n_a}.
\label{ampsb}
\end{eqnarray}

\noindent
Now, it is the initial composite state, $|S_{\bm{8}}\rangle|\bm{\mu}_a,n_a\rangle$, that has been expanded by using Clebsch-Gordan coefficients. $\bm{\nu}_{\tau}$ goes through all representations in $\bm{8}\otimes \bm{\mu}_a$, quantum numbers $n'=n_a$ due to $n_S= (0,0,0)$, and $g_{\bm{\nu}_{\tau}}$ is a coupling constant parameter.

From the previous analysis, the general expression of the decay amplitude that includes violations up to first order of the SU(3) flavor symmetry is obtained from Eqs.~(\ref{ampdecgral}) to (\ref{ampsb}) with $H_{\rm int}=H_{\rm st}+ H_{\rm ms}$:

\begin{eqnarray}
G\left[a\to b\, c\right]
& = &
{\displaystyle\sum_{\bm{\mu}_{\gamma}}}
\left(
\begin{array}{cc}
\bm{\mu}_b & \bm{\mu}_c \\
n_b & n_c
\end{array}
\right|
\left.
\begin{array}{c}
\bm{\mu}_{\gamma}\\ n_a
\end{array}
\right)
g^{0}_{\bm{\mu}_{\gamma}}\,
\delta_{\bm{\mu}_{\gamma},\bm{\mu}_a} 
\nonumber\\ 
& + &
{\displaystyle\sum_{\bm{\mu}_{\gamma}}}
\left(
\begin{array}{cc}
\bm{\mu}_b & \bm{\mu}_c \\
n_b & n_c
\end{array}
\right|
\left.
\begin{array}{c}
\bm{\mu}_{\gamma}\\ n_a
\end{array}
\right)
{\displaystyle\sum_{\bm{\nu}_{\tau}}}
\left(
\begin{array}{cc}
\bm{8} & \bm{\mu}_a \\
n_S & n_a
\end{array}
\right|
\left.
\begin{array}{c}
\bm{\nu}_{\tau} \\ n_a
\end{array}
\right)
g_{\bm{\nu}_{\tau}}
\,
\delta_{\bm{\mu}_{\gamma},\bm{\nu}_{\tau}},
\label{ampgral}
\end{eqnarray}

\noindent
first line in the right hand side corresponds to the SU(3) symmetry limit and second line to SU(3) first order breaking. Representation $\bm{\mu}_{\gamma}$ goes through all representations in $\bm{\mu}_b\otimes \bm{\mu}_c$ and $\gamma$ distinguishes repeated appearances of a representation in this expansion. Similarly, $\bm{\nu}_{\tau}$ goes through all representations in $\bm{8}\otimes \bm{\mu}_a$ and $\tau$ distinguishes repeated appearances of a representation in $\bm{8}\otimes \bm{\mu}_a$.

\section{\texorpdfstring{$\bm{D_{10}\to B_8+R_{10}}$}{D10 into B + R}}\label{sec.d10br} 

Strong decays of dibaryons in the decuplet into an octet plus a decuplet of baryons, $D_{\bm{10}}\to B_{\bm{8}}+R_{\bm{10}}$, are analyzed.

Decay amplitudes in first order SU(3) breaking will be obtained from Eq.~(\ref{ampgral}). In this case, $|a\rangle=|\bm{10},n_D\rangle \equiv |D^{\bm{10}}_{n_D}\rangle$, $|b\rangle = |\bm{8},n_B\rangle\equiv |B\rangle$, and $|c\rangle=|\bm{10},n_R\rangle\equiv |R\rangle$, so that, $\bm{\mu}_{a}=\bm{10}$, $\bm{\mu}_{b}=\bm{8}$, and $\bm{\mu}_{c}=\bm{10}$. Therefore, decompositions of composite final state ($\bm{\mu}_b\otimes \bm{\mu}_c $) and spurion-initial particle ($\bm{8}\otimes \bm{\mu}_a $) product representations are the same:

\begin{equation}
\label{8x10}
\bm{\mu}_b\otimes \bm{\mu}_c = \bm{8}\otimes \bm{\mu}_a =
\bm{8}\otimes \bm{10} = \bm{8}\oplus \bm{10}\oplus \bm{27}\oplus\bm{35},
\end{equation}

\noindent
no repeated representations are present in these expansions so that indexes $\gamma$ and $\tau$ are not necessary here.

General formula~(\ref{ampgral}) reduces in this case to:

\begin{equation}\label{gd10br}
G\left[D^{\bm{10}}_{n_D}\to B\, R\right]
=
\left(
\begin{array}{cc}
\bm{8} & \bm{10} \\
n_B & n_{R}
\end{array}
\right|
\left.
\begin{array}{c}
\bm{10} \\ n_D
\end{array}
\right)
g^0_{\bm{10}}
+
{\displaystyle\sum_{\bm{\mu}}}
\left(
\begin{array}{cc}
\bm{8} & \bm{10} \\
n_B & n_{R}
\end{array}
\right|
\left.
\begin{array}{c}
\bm{\mu} \\ n_D
\end{array}
\right)
\left(
\begin{array}{cc}
\bm{8} & \bm{10} \\
n_S & n_D
\end{array}
\right|
\left.
\begin{array}{c}
\bm{\mu} \\ n_D
\end{array}
\right)
g_{\bm{\mu}},
\end{equation}

\noindent
with $\bm{\mu} = \bm{8}, \bm{10}, \bm{27}, \bm{35}$. In exact SU(3), only one parameter, $g^0_{\bm{10}}$, determines all the strong decay amplitudes of the dibaryon decuplet $D_{\bm{10}}$ into $B_{\bm 8}+R_{\bm{10}}$. Due to SU(3) breaking, there are four extra parameters $g_{\bm{8}}$, $g_{\bm{10}}$, $g_{\bm{27}}$, and $g_{\bm{35}}$.

\subsection{\texorpdfstring{$\bm{D_{10}\to B_8+R_{10}}$}{D10 into B + R} sum rules}\label{subsec.d10brsumrules}

\begin{figure}
\centerline{\includegraphics[width=5.5in]{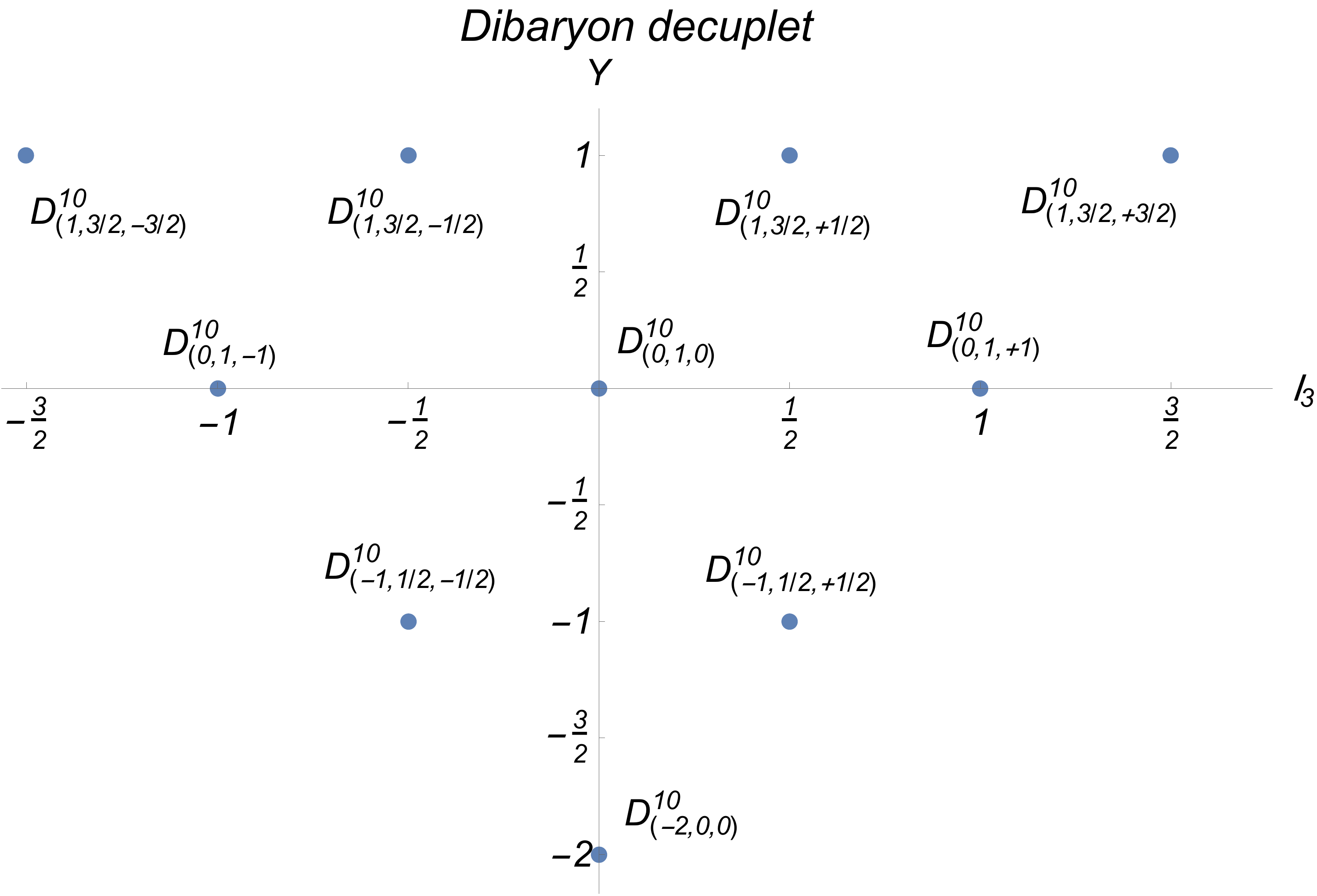}}
\caption{
\label{Dibaryons10}
$(Y,I,I_3)$ states of the dibaryon 10 multiplet $D_{\bm{10}}$.
}
\end{figure}

\begin{figure}
\centerline{\includegraphics[width=5.5in]{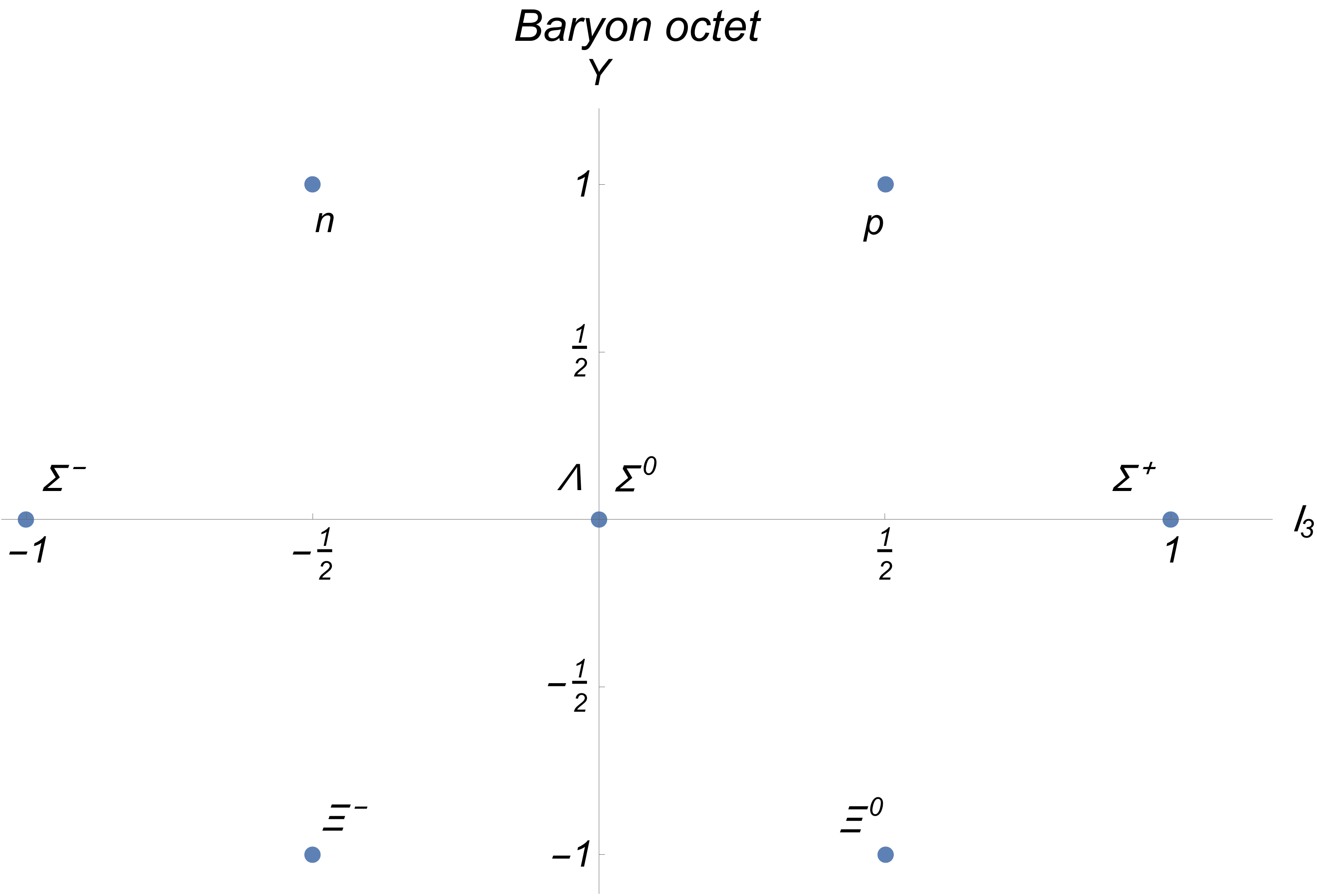}}
\caption{
\label{Baryons}
$(Y,I,I_3)$ states of the baryon octet $B_{\bm 8}$.
}
\end{figure}

\begin{figure}
\centerline{\includegraphics[width=5.5in]{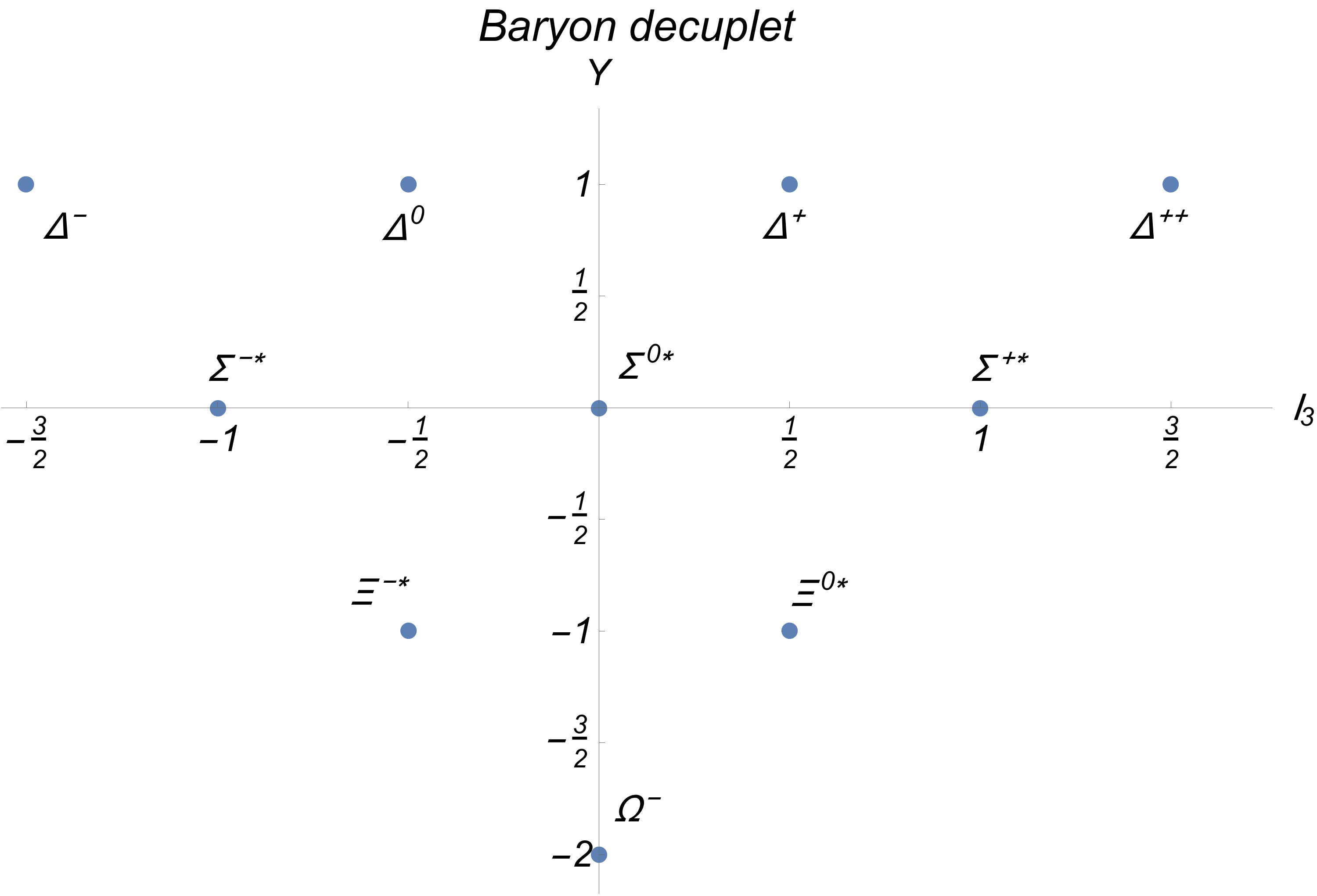}}
\caption{
\label{Resonances}
$(Y,I,I_3)$ states of the baryon decuplet $R_{\bm{10}}$.
}
\end{figure}

In this section, sum rules in broken SU(3) for the strong decays $D_{\bm{10}}\to B_{\bm{8}}+R_{\bm{10}}$ are determined. Quantum numbers of components states, $D^{\bm{10}}_{n_D}$, $B$, and $R$, of corresponding dibaryon decuplet $D_{\bm{10}}$, $J^P=(1/2)^+$ baryon octet $B_{\bm{8}}$, and $J^P=(3/2)^+$ baryon decuplet $R_{\bm{10}}$ are shown in Figs.~\ref{Dibaryons10}, \ref{Baryons}, and \ref{Resonances}, respectively.

By the use of Eq.~(\ref{gd10br}) and following sign conventions for SU(3) Clebsch-Gordan coefficients from Ref.~\cite{mcnamee64}, it is found that there are 13 linearly independent decay amplitudes, they are taken as:

\begin{equation}
G\left[D_{(1,3/2,+3/2)}^{10}\to p\,\Sigma^{+*}\right]\equiv X_1,\quad G\left[D_{(1,3/2,+3/2)}^{10}\to \Sigma^{+}\Delta^{+}\right]\equiv X_2,
\label{gd10br1}
\end{equation}

\begin{equation}
G\left[D_{(1,3/2,+3/2)}^{10}\to \Lambda\Delta^{++}\right]\equiv X_3,\quad G\left[D_{(0,1,+1)}^{10}\to p\,\Xi^{0*}\right]\equiv X_4,
\label{gd10br2}
\end{equation}

\begin{equation}
G\left[D_{(0,1,+1)}^{10}\to \Sigma^{+}\Sigma^{0*}\right]\equiv X_5,\quad G\left[D_{(0,1,+1)}^{10}\to \Lambda\Sigma^{+*}\right]\equiv X_6,
\label{gd10br3}
\end{equation}

\begin{equation}
G\left[D_{(0,1,+1)}^{10}\to \Xi^{0}\Delta^{+}\right]\equiv X_7,\quad G\left[D_{(-1,1/2,+1/2)}^{10}\to p\Omega^{-}\right]\equiv X_8,
\label{gd10br4}
\end{equation}

\begin{equation}
G\left[D_{(-1,1/2,+1/2)}^{10}\to \Sigma^{+}\Xi^{-*}\right]\equiv X_9,\quad G\left[D_{(-1,1/2,+1/2)}^{10}\to \Lambda\Xi^{0*}\right]\equiv X_{10},
\label{gd10br5}
\end{equation}

\begin{equation}
G\left[D_{(-1,1/2,+1/2)}^{10}\to \Xi^{0}\Sigma^{0*}\right]\equiv X_{11},\quad G\left[D_{(-2,0,0)}^{10}\to \Lambda\Omega^{-}\right]\equiv X_{12},
\label{gd10br6}
\end{equation}

\begin{equation}
G\left[D_{(-2,0,0)}^{10}\to \Xi^{0}\Xi^{-*}\right]\equiv X_{13}.
\label{gd10br7}
\end{equation}

\noindent
Relationships for the remaining non-zero dependent amplitudes in terms of the independent ones are given in Sec.~\ref{app:d10br} of Appendix~\ref{appendix}.

The 13 independent decay amplitudes are given in terms of the 5 parameters $g^0_{\bm{10}}$, $g_{\bm{8}}$, $g_{\bm{10}}$, $g_{\bm{27}}$, and $g_{\bm{35}}$, consequently, 8 sum rules in broken SU(3) can be established:

\begin{equation}
X_{1}-\sqrt{3}X_{4}+X_{8}=0,
\label{d10br1}
\end{equation}

\begin{equation}
-2X_{2}+\sqrt{3}X_{4}+\sqrt{6}X_{5}-2X_{8}=0,
\label{d10br2}
\end{equation}

\begin{equation}
-4X_{3}-3\sqrt{6}X_{4}+6X_{6}+6\sqrt{2}X_{8}-2X_{12}=0,
\label{d10br3}
\end{equation}

\begin{equation}
-X_{2}+\sqrt{2}X_{3}+2\sqrt{3}X_{7}+2X_{8}-\sqrt{2}X_{12}=0,
\label{d10br4}
\end{equation}

\begin{equation}
-X_{2}+\sqrt{3}X_{4}-2X_{8}+\sqrt{3}X_{9}=0,
\label{d10br5}
\end{equation}

\begin{equation}
-\sqrt{2}X_{3}-3\sqrt{3}X_{4}+6X_{8}+3\sqrt{2}X_{10}-2\sqrt{2}X_{12} = 0,
\label{d10br6}
\end{equation}

\begin{equation}
-\sqrt{2}X_{2}+2X_{3}+\sqrt{6}X_{4}+2\sqrt{3}X_{11}-2X_{12}=0,
\label{d10br7}
\end{equation}

\begin{equation}
-X_{2}+\sqrt{2}X_{3}+2\sqrt{3}X_{4}-2X_{8}-\sqrt{2}X_{12}+2X_{13}=0.
\label{d10br8}
\end{equation}

\section{\texorpdfstring{$\bm{D_{27}\to B_8+R_{10}}$}{D27 into B + R}}\label{sec.d27br} 

Strong decays of dibaryons in the $\bm{27}$-multiplet into an octet plus a decuplet of baryons, $D_{\bm{27}}\to B_{\bm{8}}+R_{\bm{10}}$, are analyzed.

SU(3) first order breaking decay amplitudes are given by Eq.~(\ref{ampgral}). In this case, $|a\rangle=|\bm{27},n_D\rangle \equiv |D^{\bm{27}}_{n_D}\rangle$, $|b\rangle = |\bm{8},n_B\rangle\equiv |B\rangle $, and $|c\rangle=|\bm{10},n_R\rangle\equiv |R\rangle$, so that, $\bm{\mu}_{a}=\bm{27}$, $\bm{\mu}_{b}=\bm{8}$, and $\bm{\mu}_{c}=\bm{10}$. Therefore,

\begin{equation}
\bm{\mu}_b\otimes \bm{\mu}_c =
\bm{8}\otimes \bm{10} =
\bm{8}\oplus \bm{10}\oplus \bm{27}\oplus\bm{35},
\end{equation}

\noindent
in this case, index $\gamma$ is not necessary because there are no repeated representations in this expansion. Also,

\begin{equation}
\bm{8}\otimes \bm{\mu}_a =
\bm{8}\otimes \bm{27} =
\bm{8}\oplus \bm{10}\oplus \bm{10^*}\oplus \bm{27}\oplus \bm{27'}\oplus \bm{35}\oplus \bm{35^*}\oplus \bm{64},
\end{equation}

\noindent
so that, index $\tau$ is necessary due to the double appearance of the $\bm{27}$.

General expression~(\ref{ampgral}) reduces to

\begin{eqnarray}
G\left[D^{\bm{27}}_{n_D}\to B\, R\right]
& = &
\left(
\begin{array}{cc}
\bm{8} & \bm{10} \\
n_B & n_R
\end{array}
\right|
\left.
\begin{array}{c}
\bm{27}\\ n_D
\end{array}
\right)
g^{0}_{\bm{27}}
\nonumber\\ 
& + &
{\displaystyle\sum_{\bm{\mu}}}
\left(
\begin{array}{cc}
\bm{8} & \bm{10} \\
n_B & n_R
\end{array}
\right|
\left.
\begin{array}{c}
\bm{\mu}\\ n_D
\end{array}
\right)
{\displaystyle\sum_{\bm{\nu}_{\tau}}}
\left(
\begin{array}{cc}
\bm{8} & \bm{27} \\
n_S & n_D
\end{array}
\right|
\left.
\begin{array}{c}
\bm{\nu}_{\tau} \\ n_D
\end{array}
\right)
g_{\bm{\nu}_{\tau}}
\,
\delta_{\bm{\mu},\bm{\nu}_{\tau}}.
\label{gd27br}
\end{eqnarray}

\noindent
with $\bm{\mu}= \bm{8}, \bm{10}, \bm{27}, \bm{35},$ and $\bm{\nu}_{\tau}= \bm{8}, \bm{10}, \bm{10^*}, \bm{27}, \bm{27'}, \bm{35}, \bm{35^*}, \bm{64}$. In exact SU(3), one parameter, $g^0_{\bm{27}}$, determines all the strong decay amplitudes of $D_{\bm{27}}$ into $B_{\bm 8}+R_{\bm{10}}$. Due to SU(3) breaking, there are five extra parameters $g_{\bm{8}}$, $g_{\bm{10}}$, $g_{\bm{27}}$, $g_{\bm{27'}}$, and $g_{\bm{35}}$.

\subsection{\texorpdfstring{$\bm{D_{27}\to B_8+R_{10}}$}{D27 into B + R} sum rules}\label{subsec.d27brsumrules} 

\begin{figure}
\centerline{\includegraphics[width=5.5in]{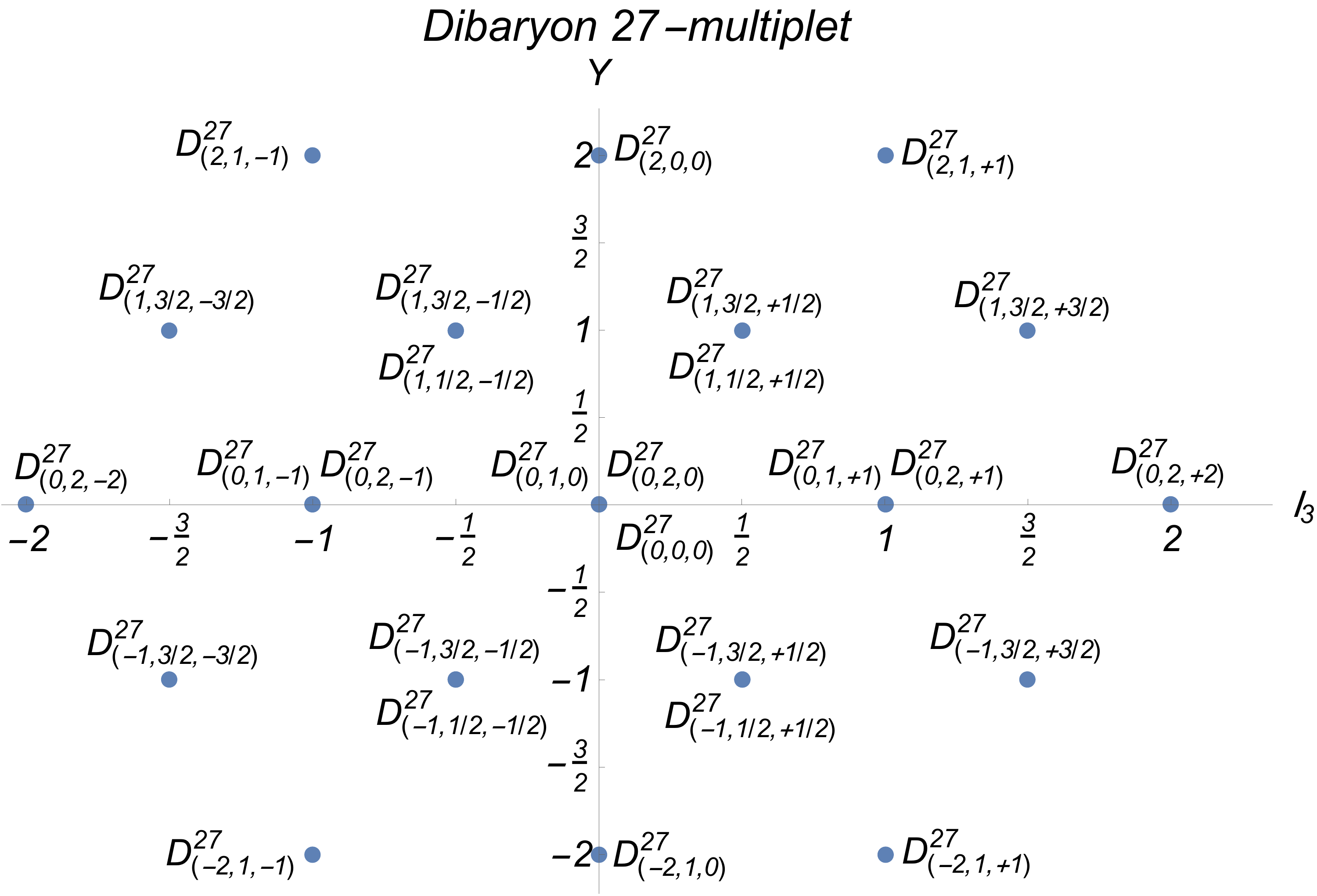}}
\caption{
\label{Dibaryons27}
$(Y,I,I_3)$ states of the dibaryon 27 multiplet $D_{\bm{27}}$.
}
\end{figure}

Sum rules for strong decay amplitudes in broken SU(3) of dibaryon $\bm{27}$-plet, $D_{\bm{27}}$ (quantum numbers of the components states, $D^{\bm{27}}_{n_D}$, are shown in Fig.~\ref{Dibaryons27}), into a final state of an ordinary baryon octet (Fig.~\ref{Baryons}) plus a baryon decuplet (Fig.~\ref{Resonances}), $D_{\bm{27}}\to B_{\bm{8}}+R_{\bm{10}}$, are determined.

From Eq.~(\ref{gd27br}) and Ref.~\cite{mcnamee64}, there are 22 linearly independent decay amplitudes that can be taken as:

\begin{equation}
G[D_{(2,1,+1)}^{27}\to p\Delta^{+}]\equiv Y_{1},
\quad
G[D_{(1,3/2,+3/2)}^{27}\to p\Sigma^{+*}]\equiv Y_{2},
\label{g27br1}
\end{equation}

\begin{equation}
G[D_{(1,3/2,+3/2)}^{27}\to\Sigma^{+}\Delta^{+}]\equiv Y_{3},
\quad
G[D_{(1,3/2,+3/2)}^{27}\to\Lambda\Delta^{++}]\equiv Y_{4},
\label{g27br2}
\end{equation}

\begin{equation}
G[D_{(1,1/2,+1/2)}^{27}\to p\Sigma^{0*}]\equiv Y_{5},
\quad
G[D_{(1,1/2,+1/2)}^{27}\to\Sigma^{+}\Delta^{0}]\equiv Y_{6},
\label{g27br3}
\end{equation}

\begin{equation}
G[D_{(0,2,+2)}^{27}\to\Sigma^{+}\Sigma^{+*}]\equiv Y_{7},
\quad
G[D_{(0,2,+2)}^{27}\to\Xi^{0}\Delta^{++}]\equiv Y_{8},
\label{g27br4}
\end{equation}

\begin{equation}
G[D_{(0,1,+1)}^{27}\to p\Xi^{0*}]\equiv Y_{9},
\quad
G[D_{(0,1,+1)}^{27}\to\Sigma^{+}\Sigma^{0*}]\equiv Y_{10},
\label{g27br5}
\end{equation}

\begin{equation}
G[D_{(0,1,+1)}^{27}\to\Lambda\Sigma^{+*}]\equiv Y_{11},
\quad
G[D_{(0,1,+1)}^{27}\to\Xi^{0}\Delta^{+}]\equiv Y_{12},
\label{g27br6}
\end{equation}

\begin{equation}
G[D_{(0,0,0)}^{27}\to p\Xi^{-*}]\equiv Y_{13},
\quad
G[D_{(0,0,0)}^{27}\to\Sigma^{+}\Sigma^{-*}]\equiv Y_{14},
\label{g27br7}
\end{equation}

\begin{equation}
G[D_{(-1,3/2,+3/2)}^{27}\to\Sigma^{+*}\Xi^{0*}]\equiv Y_{15},
\quad
G[D_{(-1,3/2,+3/2)}^{27}\to\Xi^{0}\Sigma^{+*}]\equiv Y_{16},
\label{g27br8}
\end{equation}

\begin{equation}
G[D_{(-1,1/2,+1/2)}^{27}\to p\Omega^{-}]\equiv Y_{17},
\quad
G[D_{(-1,1/2,+1/2)}^{27}\to\Sigma^{+}\Xi^{-*}]\equiv Y_{18},
\label{g27br9}
\end{equation}

\begin{equation}
G[D_{(-1,1/2,+1/2)}^{27}\to\Lambda\Xi^{0*}]\equiv Y_{19},
\quad
G[D_{(-1,1/2,+1/2)}^{27}\to\Xi^{0}\Sigma^{0*}]\equiv Y_{20},
\label{g27br10}
\end{equation}

\begin{equation}
G[D_{(-2,1,+1)}^{27}\to\Sigma^{+}\Omega^{-}]\equiv Y_{21},
\quad
G[D_{(-2,1,+1)}^{27}\to\Xi^{0}\Xi^{0*}]\equiv Y_{22}.
\label{g27br11}
\end{equation}

\noindent
Relationships for the remaining non-zero dependent amplitudes in terms of the independent ones can be found in Sec.~\ref{app:d27br} of Appendix~\ref{appendix}.

The 22 independent amplitudes are described by 6 parameters, so, 16 sum rules can be established:

\begin{equation}
Y_{1}+\sqrt{2}Y_{2}-\sqrt{2}Y_{3}-Y_{7}=0,
\label{d27br1}
\end{equation}

\begin{equation}
\sqrt{6}Y_{1}-\sqrt{3}Y_{3}-\sqrt{6}Y_{4}+\sqrt{2}Y_{8}=0,
\label{d27br2}
\end{equation}

\begin{equation}
-6\sqrt{3}Y_{1}-4\sqrt{6}Y_{2}+6\sqrt{5}Y_{5}+3\sqrt{15}Y_{9}=0,
\label{d27br3}
\end{equation}

\begin{equation}
9\sqrt{3}Y_{1}+\sqrt{6}Y_{2}-5\sqrt{6}Y_{3}-6\sqrt{5}Y_{5}+6\sqrt{10}Y_{6}+3\sqrt{30}Y_{10}=0,
\label{d27br4}
\end{equation}

\begin{equation}
3\sqrt{2}Y_{1}-2Y_{2}-2\sqrt{2}Y_{4}-\sqrt{30}Y_{5}+\sqrt{15}Y_{11}=0,
\label{d27br5}
\end{equation}

\begin{equation}
-3\sqrt{6}Y_{1}-5\sqrt{3}Y_{3}+3\sqrt{6}Y_{4}-12\sqrt{5}Y_{6}+6\sqrt{30}Y_{12}=0,
\label{d27br6}
\end{equation}

\begin{equation}
2\sqrt{3}Y_{1}-3\sqrt{5}Y_{5}+\sqrt{10}Y_{13}=0,
\label{d27br7}
\end{equation}

\begin{equation}
-4Y_{1}+\sqrt{15}Y_{5}-2\sqrt{30}Y_{6}+\sqrt{30}Y_{14}=0,
\label{d27br8}
\end{equation}

\begin{equation}
-6\sqrt{2}Y_{1}-4Y_{2}+4Y_{3}+\sqrt{30}Y_{6}-\sqrt{15}Y_{7}+3Y_{15}=0,
\label{d27br9}
\end{equation}

\begin{equation}
6\sqrt{2}Y_{1}-2Y_{2}-Y_{3}-3\sqrt{2}Y_{4}-\sqrt{30}Y_{5}+\sqrt{15}Y_{6}+3Y_{16}=0,
\label{d27br10}
\end{equation}

\begin{equation}
-3\sqrt{2}Y_{1}-Y_{2}+\sqrt{30}Y_{5}+\sqrt{5}Y_{17}=0,
\label{d27br11}
\end{equation}

\begin{equation}
21\sqrt{2}Y_{1}+2Y_{2}-5Y_{3}-5\sqrt{30}Y_{5}+8\sqrt{15}Y_{6}+3\sqrt{15}Y_{18}=0,
\label{d27br12}
\end{equation}

\begin{equation}
3Y_{1}-\sqrt{2}Y_{2}-Y_{4}-\sqrt{15}Y_{5}+\sqrt{5}Y_{19}=0,
\label{d27br13}
\end{equation}

\begin{equation}
-6\sqrt{2}Y_{1}+2\sqrt{2}Y_{2}-5Y_{3}+3\sqrt{2}Y_{4}+\sqrt{30}Y_{5}-4\sqrt{15}Y_{6}+3\sqrt{30}Y_{20}=0,
\label{d27br14}
\end{equation}

\begin{equation}
-9Y_{1}-\sqrt{2}Y_{2}+\sqrt{2}Y_{3}+2\sqrt{15}Y_{5}-\sqrt{30}Y_{6}+\sqrt{3}Y_{21}=0,
\label{d27br15}
\end{equation}

\begin{equation}
9\sqrt{6}Y_{1}-4\sqrt{3}Y_{2}+\sqrt{3}Y_{3}-3\sqrt{6}Y_{4}-6\sqrt{10}Y_{5}+6\sqrt{5}Y_{6}+3\sqrt{6}Y_{22}=0.
\label{d27br16}
\end{equation}

\section{\texorpdfstring{$\bm{D_{27}\to B_8+B_8}$}{D27 into B + B'}}\label{sec.d27bb} 

Strong decays of dibaryons in the $\bm{27}$-multiplet into two baryon octets, $D_{\bm{27}}\to B_{\bm{8}}+B_{\bm{8}}$, are analyzed.

Decay amplitudes with first order SU(3) symmetry breaking are obtained from Eq.~(\ref{ampgral}). In this case, $|a\rangle=|\bm{27},n_D\rangle \equiv |D^{\bm{27}}_{n_D}\rangle $, $|b\rangle = |\bm{8},n_B\rangle\equiv |B\rangle $, and $|c\rangle=|\bm{8},n_{B'}\rangle\equiv |B'\rangle$, so that, $\bm{\mu}_{a}=\bm{27}$, $\bm{\mu}_{b}=\bm{\mu}_{c}=\bm{8}$. For sake of clarity, a prime is placed on the baryon in second octet, but it is understood that $B$ and $B'$ are identical baryons. Here,

\begin{equation}\label{8x8}
	\bm{\mu}_b\otimes \bm{\mu}_c =
	\bm{8}\otimes \bm{8} = \bm{1}\oplus\bm{8}\oplus\bm{8'}\oplus \bm{10}\oplus\bm{10^*}\oplus\bm{27},
\end{equation}

\noindent
index $\gamma$ is necessary because the octet appears twice in this expansion. Also,

\begin{equation}
	\bm{8}\otimes \bm{\mu}_a =
	\bm{8}\otimes \bm{27} =
	\bm{8}\oplus \bm{10}\oplus \bm{10^*}\oplus \bm{27}\oplus \bm{27'}\oplus \bm{35}\oplus \bm{35^*}\oplus \bm{64}
\end{equation}

\noindent
and index $\tau$ is also necessary due to the double appearance of the $\bm{27}$.

General expression~(\ref{ampgral}) reduces to

\begin{eqnarray}
	G\left[D^{\bm{27}}_{n_D}\to B\, B'\right]
	& = &
	\left(
	\begin{array}{cc}
		\bm{8} & \bm{8} \\
		n_B & n_{B'}
	\end{array}
	\right|
	\left.
	\begin{array}{c}
		\bm{27}\\ n_D
	\end{array}
	\right)
	g^{0}_{\bm{27}}
	\nonumber\\ 
	& + &
	{\displaystyle\sum_{\bm{\mu}_{\gamma}}}
	\left(
	\begin{array}{cc}
		\bm{8} & \bm{8} \\
		n_B & n_{B'}
	\end{array}
	\right|
	\left.
	\begin{array}{c}
		\bm{\mu}_{\gamma}\\ n_D
	\end{array}
	\right)
	{\displaystyle\sum_{\bm{\nu}_{\tau}}}
	\left(
	\begin{array}{cc}
		\bm{8} & \bm{27} \\
		n_S & n_D
	\end{array}
	\right|
	\left.
	\begin{array}{c}
		\bm{\nu}_{\tau} \\ n_D
	\end{array}
	\right)
	g_{\bm{\nu}_{\tau}}
	\,
	\delta_{\bm{\mu}_{\gamma},\bm{\nu}_{\tau}},
	\label{ampdecd27bb}
\end{eqnarray}

\noindent
with $\bm{\mu}_{\gamma}= \bm{1}, \bm{8}, \bm{8'}, \bm{10}, \bm{10^*}, \bm{27}$, and $\bm{\nu}_{\tau}= \bm{8}, \bm{10}, \bm{10^*}, \bm{27}, \bm{27'}, \bm{35}, \bm{35^*}, \bm{64}$.

In the final state decomposition, Eq.~(\ref{8x8}), $\bm{8'}$ octet and $\bm{10}$ and $\bm{10^*}$ decuplets are antisymmetric under exchange of the two octet baryon states. Singlet $\bm{1}$, octet $\bm{8}$, and $\bm{27}$-plet are symmetric. There are seven parameters in Eq.~(\ref{ampdecd27bb}), three of them (all coming from SU(3) breaking) correspond to couplings with flavor-antisymmetric final state representations $\langle\bm{8'}|\bm{8}\rangle$, $\langle\bm{10}|\bm{10}\rangle$, and $\langle\bm{10^*}|\bm{10^*}\rangle$, denoted $g_{\bm{8'}}$, $g_{\bm{10}}$, and $g_{\bm{10^*}}$, respectively. Couplings with flavor-symmetric final state representations $\langle\bm{8}|\bm{8}\rangle$, $\langle\bm{27}|\bm{27}\rangle$, and $\langle\bm{27}|\bm{27'}\rangle$, introduce four additional parameters, $g^{0}_{\bm{27}}$ (from exact SU(3)) and $g_{\bm{8}}$, $g_{\bm{27}}$, and $g_{\bm{27'}}$ (from SU(3) first order breaking).

\subsection{\texorpdfstring{$\bm{D_{27}\to B_8+B_8}$}{D27 into B + B'} sum rules}\label{subsec.d27bbsumrules}

Due to the presence of two identical fermions in the final state of $D_{\bm{27}}\to B_{\bm{8}}+B_{\bm{8}}$ decays, it is important to consider symmetry properties under interchange of the two final baryons. Both cases, antisymmetric and symmetric final state, will be analyzed separately.

\subsubsection{Flavor-antisymmetric final state}\label{subsubsec.d27bba} 

Sum rules for the strong decays of the $\bm{27}$-plet of dibaryons (Fig.~\ref{Dibaryons27}) into a flavor-antisymmetric final state of two ordinary $J^P=(1/2)^+$ baryon octets (Fig.~\ref{Baryons}) are determined. As mentioned above, in this case, the corresponding decay amplitudes will be described by 3 parameters, $g_{\bm{8'}}$, $g_{\bm{10}}$, and $g_{\bm{10^*}}$, all of them from first order SU(3) symmetry breaking.

From Eq.~(\ref{ampdecd27bb}) and Ref.~\cite{mcnamee64}, there are 10 linearly independent decay amplitudes, they can be chosen as:

\begin{equation}
G[D_{(1,3/2,+3/2)}^{27}\to p\Sigma^{+'}]\equiv Z^{A}_{1},
\quad
G[D_{(1,1/2,+1/2)}^{27}\to p\Sigma^{0'}]\equiv Z^{A}_{2},
\label{g27bba1}
\end{equation}

\begin{equation}
G[D_{(1,1/2,+1/2)}^{27}\to n\Sigma^{+'}]\equiv Z^{A}_{3},
\quad
G[D_{(0,1,+1)}^{27}\to p\Xi^{0'}]\equiv Z^{A}_{4},
\label{g27bba2}
\end{equation}

\begin{equation}
G[D_{(0,1,+1)}^{27}\to\Sigma^{+}\Sigma^{0'}]\equiv Z^{A}_{5},
\quad
G[D_{(0,+1,+1)}^{27}\to\Sigma^{+}\Lambda^{'}]\equiv Z^{A}_{6},
\label{g27bba3}
\end{equation}

\begin{equation}
G[D_{(0,0,0)}^{27}\to p\Xi^{-'}]\equiv Z^{A}_{7},
\quad
G[D_{(-1,1/2,+1/2)}^{27}\to\Sigma^{+}\Xi^{-'}]\equiv Z^{A}_{8},
\label{g27bba4}
\end{equation}

\begin{equation}
G[D_{(-1,1/2,+1/2)}^{27}\to\Lambda\Xi^{0'}]\equiv Z^{A}_{9},
\quad
G[D_{(-1,3/2,+3/2)}^{27}\to\Sigma^{+}\Xi^{0'}]\equiv Z^{A}_{10}.
\label{g27bba5}
\end{equation}

\noindent
Relationships for the remaining non-zero dependent amplitudes in terms of the independent ones are given in Sec.~\ref{app:d27antybb} of Appendix~\ref{appendix}.

Since the 10 independent decay amplitudes are described by three parameters, seven SU(3) broken sum rules may be established:

\begin{equation}
4Z^{A}_{1}-\sqrt{30}Z^{A}_{2}+3\sqrt{10}Z^{A}_{3}-3\sqrt{10}Z^{A}_{4}=0,
\label{d27bba1}
\end{equation}

\begin{equation}
2Z^{A}_{1}-2\sqrt{30}Z^{A}_{2}+3\sqrt{5}Z^{A}_{5}=0,
\label{d27bba2}
\end{equation}

\begin{equation}
2Z^{A}_{1}+\sqrt{30}Z^{A}_{2}-\sqrt{10}Z^{A}_{3}+\sqrt{15}Z^{A}_{6}=0,
\label{d27bba3}
\end{equation}

\begin{equation}
3Z^{A}_{2}+\sqrt{3}Z^{A}_{3}-2\sqrt{2}Z^{A}_{7}=0,
\label{d27bba4}
\end{equation}

\begin{equation}
\sqrt{2}Z^{A}_{1}-\sqrt{15}Z^{A}_{2}-\sqrt{5}Z^{A}_{3}+\sqrt{30}Z^{A}_{8}=0,
\label{d27bba5}
\end{equation}

\begin{equation}
2Z^{A}_{1}+\sqrt{30}Z^{A}_{2}+\sqrt{10}Z^{A}_{3}-2\sqrt{10}Z^{A}_{9}=0,
\label{d27bba6}
\end{equation}

\begin{equation}
\sqrt{15}Z^{A}_{2}-\sqrt{5}Z^{A}_{3}+\sqrt{2}Z^{A}_{10}=0.
\label{d27bba7}
\end{equation}

\subsubsection{Flavor-symmetric final state}\label{subsubsec.d27bbs} 

Now, sum rules for SU(3) symmetry breaking decays of dibaryons in the $\bm{27}$ (Fig.~\ref{Dibaryons27}) into a flavor-symmetric final state of two baryon octets (Fig.~\ref{Baryons}) are calculated. As described at the beginning of this section, four coupling constants are involved in this case: $g^{0}_{\bm{27}}$ from exact SU(3), and  $g_{\bm{8}}$, $g_{\bm{27}}$, and $g_{\bm{27'}}$, from SU(3) breaking.

From Eq.~(\ref{ampdecd27bb}) and Ref.~\cite{mcnamee64}, there exist 14 linearly independent decay amplitudes:

\begin{equation}
G[D_{(2,1,+1)}^{27}\to pp']\equiv Z^{S}_{1},
\quad
G[D_{(1,3/2,+3/2)}^{27}\to p\Sigma^{+'}]\equiv Z^{S}_{2},
\label{g27bbs1}
\end{equation}

\begin{equation}
G[D_{(1,1/2,+1/2)}^{27}\to p\Sigma^{0'}]\equiv Z^{S}_{3},
\quad
G[D_{(1,1/2,+1/2)}^{27}\to p\Lambda']\equiv Z^{S}_{4},
\label{g27bbs2}
\end{equation}

\begin{equation}
G[D_{(0,2,+2)}^{27}\to\Sigma^{+}\Sigma^{+'}]\equiv Z^{S}_{5},
\quad
G[D_{(0,1,+1)}^{27}\to p\Xi^{0'}]\equiv Z^{S}_{6},
\label{g27bbs3}
\end{equation}

\begin{equation}
G[D_{(0,1,+1)}^{27}\to\Sigma^{+}\Lambda']\equiv Z^{S}_{7},
\quad
G[D_{(0,0,0)}^{27}\to p\Xi^{-'}]\equiv Z^{S}_{8},
\label{g27bbs4}
\end{equation}

\begin{equation}
G[D_{(0,0,0)}^{27}\to\Sigma^{+}\Sigma^{-'}]\equiv Z^{S}_{9},  \quad
G[D_{(0,0,0)}^{27}\to\Lambda\Lambda']\equiv Z^{S}_{10},
\label{g27bbs5}
\end{equation}

\begin{equation}
G[D_{(-1,3/2,+3/2)}^{27}\to\Sigma^{+}\Xi^{0'}]\equiv Z^{S}_{11},
\quad
G[D_{(-1,1/2,+1/2)}^{27}\to\Sigma^{+}\Xi^{-'}]\equiv Z^{S}_{12},
\label{g27bbs6}
\end{equation}

\begin{equation}
G[D_{(-1,1/2,+1/2)}^{27}\to\Lambda\Xi^{0'}]\equiv Z^{S}_{13},
\quad
G[D_{(-2,1,+1)}^{27}\to\Xi^{0}\Xi^{0'}]\equiv Z^{S}_{14},
\label{g27bbs7}
\end{equation}

\noindent
Relationships for the remaining non-zero dependent amplitudes in terms of the independent ones are given in Sec.~\ref{app:d27symbb} of Appendix~\ref{appendix}.

The 14 independent amplitudes are described by four parameters and ten SU(3) broken sum rules may be deduced:

\begin{equation}
Z^{S}_{1}-2\sqrt{2}Z^{S}_{2}+Z^{S}_{5}=0,
\label{d27bbs1}
\end{equation}

\begin{equation}
3\sqrt{3}Z^{S}_{1}-2\sqrt{6}Z^{S}_{2}+3\sqrt{5}Z^{S}_{3}-3\sqrt{15}Z^{S}_{4}+3\sqrt{15}Z^{S}_{6}=0,
\label{d27bbs2}
\end{equation}

\begin{equation}
3Z^{S}_{1}-2\sqrt{2}Z^{S}_{2}-2\sqrt{15}Z^{S}_{3}-2\sqrt{5}Z^{S}_{4}+\sqrt{30}Z^{S}_{7}=0,
\label{d27bbs3}
\end{equation}

\begin{equation}
\sqrt{3}Z^{S}_{1}-3\sqrt{5}Z^{S}_{3}-\sqrt{15}Z^{S}_{4}+2\sqrt{10}Z^{S}_{8}=0,
\label{d27bbs4}
\end{equation}

\begin{equation}
Z^{S}_{1}-4\sqrt{15}Z^{S}_{3}-2\sqrt{30}Z^{S}_{9}=0,
\label{d27bbs5}
\end{equation}

\begin{equation}
3\sqrt{3}Z^{S}_{1}-4\sqrt{15}Z^{S}_{4}+2\sqrt{10}Z^{S}_{10}=0,
\label{d27bbs6}
\end{equation}

\begin{equation}
6Z^{S}_{1}-4\sqrt{2}Z^{S}_{2}-\sqrt{15}Z^{S}_{3}-3\sqrt{5}Z^{S}_{4}+3\sqrt{2}Z^{S}_{11}=0,
\label{d27bbs7}
\end{equation}

\begin{equation}
6Z^{S}_{1}-\sqrt{2}Z^{S}_{2}-7\sqrt{15}Z^{S}_{3}-3\sqrt{5}Z^{S}_{4}+3\sqrt{30}Z^{S}_{12}=0,
\label{d27bbs8}
\end{equation}

\begin{equation}
6Z^{S}_{1}-\sqrt{2}Z^{S}_{2}-\sqrt{15}Z^{S}_{3}-5\sqrt{5}Z^{S}_{4}+2\sqrt{5}Z^{S}_{13}=0,
\label{d27bbs9}
\end{equation}

\begin{equation}
9\sqrt{3}Z^{S}_{1}-2\sqrt{6}Z^{S}_{2}-6\sqrt{5}Z^{S}_{3}-6\sqrt{15}Z^{S}_{4}+3\sqrt{3}Z^{S}_{14}=0.
\label{d27bbs10}
\end{equation}

\section{Summary and concluding remarks}\label{sec.conclusions}

If a dibaryon is made of two baryon octets, its SU(3) transformation properties can be $\bm{1}$, $\bm{8}$, $\bm{10}$, $\bm{10^*}$, or $\bm{27}$. If a dibaryon is made of one baryon octet and one baryon decuplet, then it can transform as $\bm{8}$, $\bm{10}$, $\bm{27}$, or $\bm{35}$.  In Refs.~\cite{polanco16a,polanco16b,gupta15}, sum rules for strong decays in broken SU(3) have been reported for $D_{\bm{8,\,10,\,10^*}}\to B_{\bm{8}}+B_{\bm{8}}$ and for $D_{\bm{8}}\to B_{\bm{8}}+R_{\bm{10}}$ decays.

In this work, previous studies are complemented and sum rules of strong decay amplitudes in first order SU(3) symmetry breaking are determined for dibaryon decays $D_{\bm{10,\,27}}\to B_{\bm{8}}+R_{\bm{10}}$ and $D_{\bm{27}}\to B_{\bm{8}}+B_{\bm{8}}$. Summary of results obtained and references to corresponding equations are given in Table~\ref{tablai}.

Experimentally, in each decay the width $\Gamma$ determines the decay amplitude $G$ since $\Gamma\propto|{\cal M}|^2$ (with ${\cal M}$ the Feynman invariant amplitude) and ${\cal M}\propto G$. Notice that although dibaryons in the $\bm{10}$ and $\bm{27}$ multiplets are very unlikely to be stable, some of the decay modes may be kinematically forbidden and the corresponding decay amplitude would be zero, which would simplify sum rules where it appears.

It is worth noting that with appropriate changes, sum rules determined here and in previous works~\cite{polanco16a,polanco16b,gupta15,gupta64} apply to decays of other multiquark states like tetraquarks and pentaquarks. This subject will be studied separately.

\begin{acknowledgments}

G.~S\'anchez-Col\'on and E.~N.~Polanco-Eu\'an would like to thank CONACyT (M\'exico) for partial support.

\end{acknowledgments}

\begin{appendices}

\renewcommand\thesubsection{\arabic{subsection}}
\renewcommand\thesubsubsection{(\alph{subsubsection})}

\section{Dependent decay amplitudes}\label{appendix}

Expressions for linearly dependent non-zero decay amplitudes in terms of the chosen independent ones are given. These relationships were determined by following the sign convention for SU(3) Clebsch-Gordan coefficients from Ref.~\cite{mcnamee64}.

\subsection{\texorpdfstring{$\bm{D_{10}\to B_8+R_{10}}$}{D10 into B + R}}\label{app:d10br}

\begin{equation}
G[D_{(1,3/2,+3/2)}^{10}\to \Sigma^{0}\Delta^{++}]=-\sqrt{\frac{3}{2}}X_{2},
\label{gd10brothers1}
\end{equation}

\begin{equation}
G[D_{(1,3/2,+1/2)}^{10}\to p\Sigma^{0*}]=\sqrt{\frac{2}{3}}X_{1},
\quad
G[D_{(1,3/2,+1/2)}^{10}\to n\Sigma^{+*}]=\frac{1}{\sqrt{3}}X_{1},
\label{gd10brothers2}
\end{equation}

\begin{equation}
G[D_{(1,3/2,+1/2)}^{10}\to \Sigma^{+}\Delta^{0}]=\frac{2}{\sqrt{3}}X_{2},
\quad
G[D_{(1,3/2,+1/2)}^{10}\to \Sigma^{0}\Delta^{+}]=-\frac{1}{\sqrt{6}}X_{2},
\label{gd10brothers3}
\end{equation}

\begin{equation}
G[D_{(1,3/2,+1/2)}^{10}\to \Sigma^{-}\Delta^{++}]=-X_{2},
\quad
G[D_{(1,3/2,+1/2)}^{10}\to \Lambda\Delta^{+}]=X_{3},
\label{gd10brothers4}
\end{equation}

\begin{equation}
G[D_{(1,3/2,-1/2)}^{10}\to p\Sigma^{-*}]=\frac{1}{\sqrt{3}}X_{1},
\quad
G[D_{(1,3/2,-1/2)}^{10}\to n\Sigma^{0*}]=\sqrt{\frac{2}{3}}X_{1},
\label{gd10brothers5}
\end{equation}

\begin{equation}
G[D_{(1,3/2,-1/2)}^{10}\to \Sigma^{+}\Delta^{-}]=X_{2},
\quad
G[D_{(1,3/2,-1/2)}^{10}\to \Sigma^{0}\Delta^{0}]=\frac{1}{\sqrt{6}}X_{2},
\label{gd10brothers6}
\end{equation}

\begin{equation}
G[D_{(1,3/2,-1/2)}^{10}\to \Sigma^{-}\Delta^{+}]=-\frac{2}{\sqrt{3}}X_{2},
\quad
G[D_{(1,3/2,-1/2)}^{10}\to \Lambda\Delta^{0}]=X_{3},
\label{gd10brothers7}
\end{equation}

\begin{equation}
G[D_{(1,3/2,-3/2)}^{10}\to n\Sigma^{-*}]=X_{1},
\quad
G[D_{(1,3/2,-3/2)}^{10}\to \Sigma^{0}\Delta^{-}]=\sqrt{\frac{3}{2}}X_{2},
\label{gd10brothers8}
\end{equation}

\begin{equation}
G[D_{(1,3/2,-3/2)}^{10}\to \Sigma^{-}\Delta^{0}]=-X_{2},
\quad
G[D_{(1,3/2,-3/2)}^{10}\to \Lambda\Delta^{-}]=X_{3},
\label{gd10brothers9}
\end{equation}

\begin{equation}
G[D_{(0,1,+1)}^{10}\to \Sigma^{0}\Sigma^{+*}]=-X_{5},
\quad
G[D_{(0,1,+1)}^{10}\to \Xi^{-}\Delta^{++}]=-\sqrt{3}X_{7},
\label{gd10brothers10}
\end{equation}

\begin{equation}
G[D_{(0,1,0)}^{10}\to p\Xi^{-*}]=\frac{1}{\sqrt{2}}X_{4},
\quad
G[D_{(0,1,0)}^{10}\to n\Xi^{0*}]=\frac{1}{\sqrt{2}}X_{4},
\label{gd10brothers11}
\end{equation}

\begin{equation}
G[D_{(0,1,0)}^{10}\to \Sigma^{+}\Sigma^{-*}]=X_{5},
\quad
G[D_{(0,1,0)}^{10}\to \Sigma^{-}\Sigma^{+*}]=-X_{5},
\label{gd10brothers12}
\end{equation}

\begin{equation}
G[D_{(0,1,0)}^{10}\to \Lambda\Sigma^{0*}]=X_{6},
\quad
G[D_{(0,1,0)}^{10}\to \Xi^{0}\Delta^{0}]=\sqrt{2}X_{7},
\label{gd10brothers13}
\end{equation}

\begin{equation}
G[D_{(0,1,0)}^{10}\to \Xi^{-}\Delta^{+}]=-\sqrt{2}X_{7},
\label{gd10brothers14}
\end{equation}

\begin{equation}
G[D_{(0,1,-1)}^{10}\to n\Xi^{-*}]=X_{4},
\quad
G[D_{(0,1,-1)}^{10}\to \Sigma^{0}\Sigma^{-*}]=X_{5},
\label{gd10brothers15}
\end{equation}

\begin{equation}
G[D_{(0,1,-1)}^{10}\to \Sigma^{-}\Sigma^{0*}]=-X_{5},
\quad
G[D_{(0,1,-1)}^{10}\to \Lambda\Sigma^{-*}]=X_{6},
\label{gd10brothers16}
\end{equation}

\begin{equation}
G[D_{(0,1,-1)}^{10}\to \Xi^{0}\Delta^{-}]=\sqrt{3}X_{7},
\quad
G[D_{(0,1,-1)}^{10}\to \Xi^{-}\Delta^{0}]=-X_{7},
\label{gd10brothers17}
\end{equation}

\begin{equation}
G[D_{(-1,1/2,+1/2)}^{10}\to \Sigma^{0}\Xi^{0*}]=-\frac{1}{\sqrt{2}}X_{9},
\quad
G[D_{(-1,1/2,+1/2)}^{10}\to \Xi^{-}\Sigma^{+*}]=-\sqrt{2}X_{11},
\label{gd10brothers18}
\end{equation}

\begin{equation}
G[D_{(-1,1/2,-1/2)}^{10}\to n\Omega^{-}]=X_{8},
\quad
G[D_{(-1,1/2,-1/2)}^{10}\to \Sigma^{0}\Xi^{-*}]=\frac{1}{\sqrt{2}}X_{9},
\label{gd10brothers19}
\end{equation}

\begin{equation}
G[D_{(-1,1/2,-1/2)}^{10}\to \Sigma^{-}\Xi^{0*}]=-X_{9},
\quad
G[D_{(-1,1/2,-1/2)}^{10}\to \Lambda\Xi^{-*}]=X_{10},
\label{gd10brothers20}
\end{equation}

\begin{equation}
G[D_{(-1,1/2,-1/2)}^{10}\to \Xi^{0}\Sigma^{-*}]=\sqrt{2}X_{11},
\quad
G[D_{(-1,1/2,-1/2)}^{10}\to \Xi^{-}\Sigma^{0*}]=-X_{11},
\label{gd10brothers21}
\end{equation}

\begin{equation}
G[D_{(-2,0,0)}^{10}\to \Xi^{-}\Xi^{0*}]=-X_{13}.
\label{gd10brothers22}
\end{equation}

\subsection{\texorpdfstring{$\bm{D_{27}\to B_8+R_{10}}$}{D27 into sym B + R}}\label{app:d27br}

\begin{equation}
G[D_{(2,1,+1)}^{27}\to n\Delta^{++}]=-\sqrt{3}Y_{1},
\quad
G[D_{(2,1,0)}^{27}\to p\Delta^{0}]=\sqrt{2}Y_{1},
\label{d27bro1}
\end{equation}

\begin{equation}
G[D_{(2,1,0)}^{27}\to n\Delta^{+}]=-\sqrt{2}Y_{1},
\quad
G[D_{(2,1,-1)}^{27}\to p\Delta^{-}]=\sqrt{3}Y_{1},
\label{d27bro2}
\end{equation}

\begin{equation}
G[D_{(2,1,-1)}^{27}\to n\Delta^{0}]=-Y_{1},
\label{d27bro3}
\end{equation}

\begin{equation}
G[D_{(1,3/2,+3/2)}^{27}\to \Sigma^{0}\Delta^{++}]=-\sqrt{\frac{3}{2}}Y_{2},
\quad
G[D_{(1,3/2,+1/2)}^{27}\to p\Sigma^{0*}]=\sqrt{\frac{2}{3}}Y_{2},
\label{d27bro4}
\end{equation}

\begin{equation}
G[D_{(1,3/2,+1/2)}^{27}\to n\Sigma^{+*}]=\frac{1}{\sqrt{3}}Y_{2},
\quad
G[D_{(1,3/2,+1/2)}^{27}\to \Sigma^{+}\Delta^{0}]=\frac{2}{\sqrt{3}}Y_{3},
\label{d27bro5}
\end{equation}

\begin{equation}
G[D_{(1,3/2,+1/2)}^{27}\to \Sigma^{0}\Delta^{+}]=-\frac{1}{\sqrt{6}}Y_{3},
\quad
G[D_{(1,3/2,+1/2)}^{27}\to \Sigma^{-}\Delta^{++}]=-Y_{3},
\label{d27bro6}
\end{equation}

\begin{equation}
G[D_{(1,3/2,+1/2)}^{27}\to \Lambda^{0}\Delta^{+}]=Y_{4},
\quad
G[D_{(1,3/2,-1/2)}^{27}\to p\Sigma^{-*}]=\frac{1}{\sqrt{3}}Y_{2},
\label{d27bro7}
\end{equation}

\begin{equation}
G[D_{(1,3/2,-1/2)}^{27}\to n\Sigma^{0*}]=\sqrt{\frac{2}{3}}Y_{2},
\quad
G[D_{(1,3/2,-1/2)}^{27}\to \Sigma^{+}\Delta^{-}]=Y_{3},
\label{d27bro8}
\end{equation}

\begin{equation}
G[D_{(1,3/2,-1/2)}^{27}\to \Sigma^{0}\Delta^{0}]=\frac{1}{\sqrt{6}}Y_{2},
\quad
G[D_{(1,3/2,-1/2)}^{27}\to \Sigma^{-}\Delta^{+}]=-\frac{2}{\sqrt{3}}Y_{3},
\label{d27bro9}
\end{equation}

\begin{equation}
G[D_{(1,3/2,-1/2)}^{27}\to \Lambda\Delta^{0}]=Y_{4},
\quad
G[D_{(1,3/2,-3/2)}^{27}\to n\Sigma^{-*}]=Y_{2},
\label{d27bro10}
\end{equation}

\begin{equation}
G[D_{(1,3/2,-3/2)}^{27}\to \Sigma^{0}\Delta^{-}]=\sqrt{\frac{3}{2}}Y_{3},
\quad
G[D_{(1,3/2,-3/2)}^{27}\to \Sigma^{-}\Delta^{0}]=-Y_{3},
\label{d27bro11}
\end{equation}

\begin{equation}
G[D_{(1,3/2,-3/2)}^{27}\to \Lambda\Delta^{-}]=Y_{4},
\label{d27bro12}
\end{equation}

\begin{equation}
G[D_{(1,1/2,+1/2)}^{27}\to n\Sigma^{+*}]=-\sqrt{2}Y_{5},
\quad
G[D_{(1,1/2,+1/2)}^{27}\to \Sigma^{0}\Delta^{+}]=-\sqrt{2}Y_{6},
\label{d27bro13}
\end{equation}

\begin{equation}
G[D_{(1,1/2,+1/2)}^{27}\to \Sigma^{-}\Delta^{++}]=\sqrt{3}Y_{6},
\quad
G[D_{(1,1/2,-1/2)}^{27}\to p\Sigma^{-*}]=\sqrt{2}Y_{5},
\label{d27bro14}
\end{equation}

\begin{equation}
G[D_{(1,1/2,-1/2)}^{27}\to n\Sigma^{0*}]=-Y_{5},
\quad
G[D_{(1,1/2,-1/2)}^{27}\to \Sigma^{+}\Delta^{-}]=\sqrt{3}Y_{6},
\label{d27bro15}
\end{equation}

\begin{equation}
G[D_{(1,1/2,-1/2)}^{27}\to \Sigma^{0}\Delta^{0}]=-\sqrt{2}Y_{6},
\quad
G[D_{(1,1/2,-1/2)}^{27}\to \Sigma^{-}\Delta^{+}]=Y_{6},
\label{d27bro16}
\end{equation}

\begin{equation}
G[D_{(0,2,+1)}^{27}\to \Sigma^{+}\Sigma^{0*}]=\frac{1}{\sqrt{2}}Y_{7},
\quad
G[D_{(0,2,+1)}^{27}\to \Sigma^{0}\Sigma^{+*}]=\frac{1}{\sqrt{2}}Y_{7},
\label{d27bro17}
\end{equation}

\begin{equation}
G[D_{(0,2,+1)}^{27}\to \Xi^{0}\Delta^{+}]=\frac{\sqrt{3}}{2}Y_{8},
\quad
G[D_{(0,2,+1)}^{27}\to \Xi^{-}\Delta^{++}]=\frac{1}{2}Y_{8},
\label{d27bro18}
\end{equation}

\begin{equation}
G[D_{(0,2,-0)}^{27}\to \Sigma^{+}\Sigma^{-*}]=\frac{1}{\sqrt{6}}Y_{7},
\quad
G[D_{(0,2,0)}^{27}\to \Sigma^{0}\Sigma^{0*}]=\sqrt{\frac{2}{3}}Y_{7},
\label{d27bro19}
\end{equation}

\begin{equation}
G[D_{(0,2,0)}^{27}\to \Sigma^{-}\Sigma^{+*}]=\frac{1}{\sqrt{6}}Y_{7},
\quad
G[D_{(0,2,0)}^{27}\to \Xi^{0}\Delta^{0}]=\frac{1}{\sqrt{2}}Y_{8},
\label{d27bro20}
\end{equation}

\begin{equation}
G[D_{(0,2,0)}^{27}\to \Xi^{-}\Delta^{+}]=\frac{1}{\sqrt{2}}Y_{8},
\quad
G[D_{(0,2,-1)}^{27}\to \Sigma^{0}\Sigma^{-*}]=\frac{1}{\sqrt{2}}Y_{7},
\label{d27bro21}
\end{equation}

\begin{equation} 
G[D_{(0,2,-1)}^{27}\to \Sigma^{-}\Sigma^{0*}]=\frac{1}{\sqrt{2}}Y_{7},
\quad
G[D_{(0,2,-1)}^{27}\to \Xi^{0}\Delta^{-}]=\frac{1}{2}Y_{8},
\label{d27bro22}
\end{equation}

\begin{equation}
G[D_{(0,2,-1)}^{27}\to \Xi^{-}\Delta^{0}]=\frac{\sqrt{3}}{2}Y_{8},
\quad
G[D_{(0,2,-2)}^{27}\to \Sigma^{-}\Sigma^{-*}]=Y_{7},
\label{d27bro23}
\end{equation}

\begin{equation}
G[D_{(0,2,-1)}^{27}\to \Xi^{-}\Delta^{-}]=Y_{8},
\label{d27bro24}
\end{equation}

\begin{equation}
G[D_{(0,1,+1)}^{27}\to \Sigma^{0}\Sigma^{+*}]=-Y_{10},
\quad
G[D_{(0,1,+1)}^{27}\to \Xi^{-}\Delta^{++}]=-\sqrt{3}Y_{12},
\label{d27bro25}
\end{equation}

\begin{equation}
G[D_{(0,1,0)}^{27}\to p\Xi^{-*}]=\frac{1}{\sqrt{2}}Y_{9},
\quad
G[D_{(0,1,0)}^{27}\to n\Xi^{0*}]=\frac{1}{\sqrt{2}}Y_{9},
\label{d27bro26}
\end{equation}

\begin{equation}
G[D_{(0,1,0)}^{27}\to \Sigma^{+}\Sigma^{-*}]=Y_{10},
\quad
G[D_{(0,1,0)}^{27}\to \Sigma^{-}\Sigma^{+*}]=-Y_{10},
\label{d27bro27}
\end{equation}

\begin{equation}
G[D_{(0,1,0)}^{27}\to \Lambda\Sigma^{0*}]=Y_{11},
\quad
G[D_{(0,1,0)}^{27}\to \Xi^{0}\Delta^{0}]=\sqrt{2}Y_{12},
\label{d27bro28}
\end{equation}

\begin{equation}
G[D_{(0,1,0)}^{27}\to \Xi^{-}\Delta^{+}]=-\sqrt{2}Y_{12},
\quad
G[D_{(0,1,-1)}^{27}\to n\Xi^{-}]=Y_{9},
\label{d27bro29}
\end{equation}

\begin{equation}
G[D_{(0,1,-1)}^{27}\to \Sigma^{0}\Sigma^{-*}]=Y_{10},
\quad
G[D_{(0,1,-1)}^{27}\to \Sigma^{-}\Sigma^{0*}]=-Y_{10},
\label{d27bro30}
\end{equation}

\begin{equation}
G[D_{(0,1,-1)}^{27}\to \Lambda\Sigma^{-*}]=Y_{11},
\quad
G[D_{(0,1,-1)}^{27}\to \Xi^{0}\Delta^{-}]=\sqrt{3}Y_{12},
\label{d27bro31}
\end{equation}

\begin{equation}
G[D_{(0,1,-1)}^{27}\to \Xi^{-}\Delta^{0}]=-Y_{12},
\label{d27bro32}
\end{equation}

\begin{equation}
G[D_{(0,0,0)}^{27}\to n\Xi^{0*}]=-Y_{13}, \\
G[D_{(0,0,0)}^{27}\to \Sigma^{0}\Sigma^{0}]=-Y_{14},
\label{d27bro33}
\end{equation}

\begin{equation}
G[D_{(0,0,0)}^{27}\to \Sigma^{-}\Sigma^{+}]=Y_{14},
\label{d27bro34}
\end{equation}

\begin{equation}
G[D_{(-1,3/2,-1/2)}^{27}\to \Sigma^{+}\Xi^{-*}]=\frac{1}{\sqrt{3}}Y_{15},
\quad
G[D_{(-1,3/2,+1/2)}^{27}\to \Sigma^{0}\Xi^{0*}]=\sqrt{\frac{2}{3}}Y_{15},
\label{d27bro35}
\end{equation}

\begin{equation}
G[D_{(-1,3/2,-1/2)}^{27}\to \Xi^{0}\Sigma^{0*}]=\sqrt{\frac{2}{3}}Y_{16},
\quad
G[D_{(-1,3/2,+1/2)}^{27}\to \Xi^{-}\Sigma^{+*}]=\frac{1}{\sqrt{3}}Y_{16},
\label{d27bro36}
\end{equation}

\begin{equation}
G[D_{(-1,3/2,-1/2)}^{27}\to \Sigma^{0}\Xi^{-*}]=\sqrt{\frac{2}{3}}Y_{15},
\quad
G[D_{(-1,3/2,-1/2)}^{27}\to \Sigma^{-}\Xi^{0*}]=\frac{1}{\sqrt{3}}Y_{15},
\label{d27bro37}
\end{equation}

\begin{equation}
G[D_{(-1,3/2,-1/2)}^{27}\to \Xi^{0}\Sigma^{-*}]=\frac{1}{\sqrt{3}}Y_{16},
\quad
G[D_{(-1,3/2,-1/2)}^{27}\to \Xi^{-}\Sigma^{0*}]=\sqrt{\frac{2}{3}}Y_{16},
\label{d27bro38}
\end{equation}

\begin{equation}
G[D_{(-1,3/2,-1/2)}^{27}\to \Sigma^{-}\Xi^{-*}]=Y_{15},
\quad
G[D_{(-1,3/2,-1/2)}^{27}\to \Xi^{-}\Sigma^{-*}]=Y_{16},
\label{d27bro39}
\end{equation}

\begin{equation}
G[D_{(-1,1/2,+1/2)}^{27}\to \Sigma^{-}\Xi^{-*}]=-\frac{1}{\sqrt{2}}Y_{18},
\quad
G[D_{(-1,1/2,+1/2)}^{27}\to \Sigma^{-}\Xi^{-*}]=-\sqrt{2}Y_{20},
\label{d27bro40}
\end{equation}

\begin{equation}
G[D_{(-1,1/2,-1/2)}^{27}\to n\Omega^{-}]=Y_{17},
\quad
G[D_{(-1,1/2,-1/2)}^{27}\to \Sigma^{0}\Xi^{-*}]=\frac{1}{\sqrt{2}}Y_{18},
\label{d27bro41}
\end{equation}

\begin{equation}
G[D_{(-1,1/2,-1/2)}^{27}\to \Sigma^{-}\Xi^{0*}]=-Y_{18},
\quad
G[D_{(-1,1/2,-1/2)}^{27}\to \Lambda\Xi^{-*}]=Y_{19},
\label{d27bro42}
\end{equation}

\begin{equation}
G[D_{(-1,1/2,-1/2)}^{27}\to\Xi^{0}\Sigma^{-*}]=\sqrt{2}Y_{20},
\quad
G[D_{(-1,1/2,-1/2)}^{27}\to \Xi^{-}\Sigma^{0*}]=-Y_{20},
\label{d27bro43}
\end{equation}

\begin{equation}
G[D_{(-2,1,0)}^{27}\to \Sigma^{0}\Omega^{-}]=Y_{21},
\quad
G[D_{(-2,1,0)}^{27}\to \Xi^{0}\Xi^{-*}]=\frac{1}{\sqrt{2}}Y_{22},
\label{d27bro44}
\end{equation}

\begin{equation}
G[D_{(-2,1,0)}^{27}\to \Xi^{-}\Xi^{0*}]=\frac{1}{\sqrt{2}}Y_{22},
\quad
G[D_{(-2,1,-1)}^{27}\to \Sigma^{-}\Omega^{-}]=Y_{21},
\label{d27bro45}
\end{equation}

\begin{equation}
G[D_{(-2,1,-1)}^{27}\to \Xi^{-}\Xi^{-*}]=Y_{22}.
\label{d27bro46}
\end{equation}

\subsection{\texorpdfstring{$\bm{D_{27}\to B_8+B_8}$}{D27 into B + B}}\label{app:d27bb}

\subsubsection{Antisymmetric final state}\label{app:d27antybb}

\begin{equation}
G[D_{(1,3/2,+1/2)}^{27}\to p\Sigma^{0'}]=\sqrt{\frac{2}{3}}Z_{1}^{A},
\quad
G[D_{(1,3/2,+1/2)}^{27}\to n\Sigma^{+'}]=\frac{1}{\sqrt{3}}Z_{1}^{A},
\end{equation}

\begin{equation}
G[D_{(1,3/2,-1/2)}^{27}\to p\Sigma^{-'}]=\frac{1}{\sqrt{3}}Z_{1}^{A},
\quad
G[D_{(1,3/2,-1/2)}^{27}\to n\Sigma^{0'}]=\sqrt{\frac{2}{3}}Z_{1}^{A},
\end{equation}

\begin{equation}
G[D_{(1,3/2,-3/2)}^{27}\to n\Sigma^{-'}]=Z_{1}^{A},
\end{equation}

\begin{equation}
G[D_{(1,1/2,+1/2)}^{27}\to n\Sigma^{+'}]=-\sqrt{2}Z_{2}^{A},
\quad
G[D_{(1,1/2,-1/2)}^{27}\to p\Sigma^{-'}]=\sqrt{2}Z_{2}^{A},
\end{equation}

\begin{equation}
G[D_{(1,1/2,-1/2)}^{27}\to n\Sigma^{0'}]=-Z_{2}^{A},
\quad
G[D_{(1,1/2,-1/2)}^{27}\to n\Lambda']=Z_{3}^{A},
\end{equation}

\begin{equation}
G[D_{(0,1,0)}^{27}\to p\Xi^{-\text{'}}]=\frac{1}{\sqrt{2}}Z_{4}^{A},
\quad
G[D_{(0,1,0)}^{27}\to n\Xi^{0\text{'}}]=\frac{1}{\sqrt{2}}Z_{4}^{A},
\end{equation}

\begin{equation}
G[D_{(0,1,0)}^{27}\to\Sigma^{+}\Sigma^{-'}]=Z_{5}^{A},
\quad
G[D_{(0,1,0)}^{27}\to\Sigma^{0}\Lambda']=Z_{6}^{A},
\end{equation}

\begin{equation}
G[D_{(0,1,-1)}^{27}\to n\Xi^{-\text{'}}]=Z_{4}^{A},
\quad
G[D_{(0,1,-1)}^{27}\to\Sigma^{-}\Sigma^{0'}]=-Z_{5}^{A},
\end{equation}

\begin{equation}
G[D_{(0,1,-1)}^{27}\to\Sigma^{-}\Lambda']=Z_{6}^{A},
\end{equation}

\begin{equation}
G[D_{(0,0,0)}^{27}\to n\Xi^{0\text{'}}]=-Z_{7}^{A},
\end{equation}

\begin{equation}
G[D_{(-1,1/2,+1/2)}^{27}\to\Sigma^{0}\Xi^{0\text{'}}]=-\frac{1}{\sqrt{2}}Z_{8}^{A},
\quad
G[D_{(-1,1/2,-1/2)}^{27}\to\Sigma^{0}\Xi^{-\text{'}}]=\frac{1}{\sqrt{2}}Z_{8}^{A},
\end{equation}

\begin{equation}
G[D_{(-1,1/2,-1/2)}^{27}\to\Sigma^{-}\Xi^{0\text{'}}]=-Z_{8}^{A},
\quad
G[D_{(-1,1/2,-1/2)}^{27}\to\Lambda\Xi^{-'}]=Z_{9}^{A},
\end{equation}

\begin{equation}
G[D_{(-1,3/2,+1/2)}^{27}\to\Sigma^{+}\Xi^{-'}]=\frac{1}{\sqrt{3}}Z_{10}^{A},
\quad
G[D_{(-1,3/2,+1/2)}^{27}\to\Sigma^{0}\Xi^{0'}]=\sqrt{\frac{2}{3}}Z_{10}^{A},
\end{equation}

\begin{equation}
G[D_{(-1,3/2,-1/2)}^{27}\to\Sigma^{0}\Xi^{-'}]=\sqrt{\frac{2}{3}}Z_{10}^{A},
\quad
G[D_{(-1,3/2,-1/2)}^{27}\to\Sigma^{-}\Xi^{0'}]=\frac{1}{\sqrt{3}}Z_{10}^{A},
\end{equation}

\begin{equation}
G[D_{(-1,3/2,-3/2)}^{27}\to\Sigma^{-}\Xi^{-'}]=Z_{10}^{A},
\end{equation}

\noindent
with identical relationships for $G\left[D^{\bm{27}}_{(Y,I,I_3)}\to B'B\right] = -G\left[D^{\bm{27}}_{(Y,I,I_3)}\to BB'\right]$.

\subsubsection{Symmetric final state}\label{app:d27symbb}

\begin{equation}
G[D_{(2,1,0)}^{27}\to pn']=\frac{1}{\sqrt{2}}Z_{1}^{S},
\quad
G[D_{(2,1,-1)}^{27}\to nn']=Z_{1}^{S},
\end{equation}

\begin{equation}
G[D_{(1,3/2,+1/2)}^{27}\to p\Sigma^{0'}]=\sqrt{\frac{2}{3}}Z_{2}^{S},
\quad
G[D_{(1,3/2,+1/2)}^{27}\to n\Sigma^{+'}]=\frac{1}{\sqrt{3}}Z_{2}^{S},
\end{equation}

\begin{equation}
G[D_{(1,3/2,-1/2)}^{27}\to p\Sigma^{-'}]=\frac{1}{\sqrt{3}}Z_{2}^{S},
\quad
G[D_{(1,3/2,-1/2)}^{27}\to n\Sigma^{0'}]=\sqrt{\frac{2}{3}}Z_{2}^{S},
\end{equation}

\begin{equation}
G[D_{(1,3/2,-3/2)}^{27}\to n\Sigma^{-'}]=Z_{2}^{S},
\end{equation}

\begin{equation}
G[D_{(1,1/2,+1/2)}^{27}\to n\Sigma^{+'}]=-\sqrt{2}Z_{3}^{S},
\quad
G[D_{(1,1/2,-1/2)}^{27}\to p\Sigma^{-'}]=\sqrt{2}Z_{3}^{S},
\end{equation}

\begin{equation}
G[D_{(1,1/2,-1/2)}^{27}\to n\Sigma^{0'}]=-Z_{3}^{S},
\quad
G[D_{(1,1/2,-1/2)}^{27}\to n\Lambda']=Z_{4}^{S},
\end{equation}

\begin{equation}
G[D_{(0,2,+1)}^{27}\to\Sigma^{+}\Sigma^{0'}]=\frac{1}{\sqrt{2}}Z_{5}^{S},
\quad
G[D_{(0,2,0)}^{27}\to\Sigma^{+}\Sigma^{-'}]=\frac{1}{\sqrt{6}}Z_{5}^{S},
\end{equation}

\begin{equation}
G[D_{(0,2,0)}^{27}\to\Sigma^{0}\Sigma^{0'}]=\sqrt{\frac{2}{3}}Z_{5}^{S},
\quad
G[D_{(0,2,-1)}^{27}\to\Sigma^{0}\Sigma^{-'}]=\frac{1}{\sqrt{2}}Z_{5}^{S},
\end{equation}

\begin{equation}
G[D_{(0,2,-2)}^{27}\to\Sigma^{-}\Sigma^{-'}]=Z_{5}^{S},
\end{equation}

\begin{equation}
G[D_{(0,1,0)}^{27}\to p\Xi^{-'}]=\frac{1}{\sqrt{2}}Z_{6}^{S},
\quad
G[D_{(0,1,0)}^{27}\to n\Xi^{0'}]=\frac{1}{\sqrt{2}}Z_{6}^{S},
\end{equation}

\begin{equation}
G[D_{(0,1,0)}^{27}\to\Sigma^{0}\Lambda']=Z_{7}^{S},
\quad
G[D_{(0,1,-1)}^{27}\to n\Xi^{-'}]=Z_{6}^{S},
\end{equation}

\begin{equation}
G[D_{(0,1,-1)}^{27}\to\Sigma^{-}\Lambda']=Z_{7}^{S},
\end{equation}

\begin{equation}
G[D_{(0,0,0)}^{27}\to n\Xi^{0'}]=-Z_{8}^{S},
\quad
G[D_{(0,0,0)}^{27}\to\Sigma^{0}\Sigma^{0'}]=-Z_{9}^{S},
\end{equation}

\begin{equation}
G[D_{(-1,3/2,+1/2)}^{27}\to\Sigma^{+}\Xi^{-'}]=\frac{1}{\sqrt{3}}Z_{11}^{S},
\quad
G[D_{(-1,3/2,+1/2)}^{27}\to\Sigma^{0}\Xi^{0'}]=\sqrt{\frac{2}{3}}Z_{11}^{S},
\end{equation}

\begin{equation}
G[D_{(-1,3/2,-1/2)}^{27}\to\Sigma^{0}\Xi^{-'}]=\sqrt{\frac{2}{3}}Z_{11}^{S},
\quad
G[D_{(-1,3/2,-1/2)}^{27}\to\Sigma^{-}\Xi^{0'}]=\frac{1}{\sqrt{3}}Z_{11}^{S},
\end{equation}

\begin{equation}
G[D_{(-1,1/2,+1/2)}^{27}\to\Sigma^{0}\Xi^{0'}]=-\frac{1}{\sqrt{2}}Z_{12}^{S}, 
\quad
G[D_{(-1,1/2,-1/2)}^{27}\to\Sigma^{0}\Xi^{-'}]=\frac{1}{\sqrt{2}}Z_{12}^{S},
\end{equation}

\begin{equation}
G[D_{(-1,1/2,-1/2)}^{27}\to\Sigma^{-}\Xi^{0'}]=-Z_{12}^{S},
\quad
G[D_{(-1,1/2,-1/2)}^{27}\to\Lambda\Xi]=Z_{13}^{S},
\end{equation}

\begin{equation}
G[D_{(-1,3/2,-3/2)}^{27}\to\Sigma^{-}\Xi^{-'}]=Z_{11}^{S},
\end{equation}

\begin{equation}
G[D_{(-2,1,0)}^{27}\to\Xi^{0}\Xi^{-'}]=\frac{1}{\sqrt{2}}Z_{14}^{S},
\quad
G[D_{(-2,1,-1)}^{27}\to\Xi^{-}\Xi^{-'}]=Z_{14}^{S},
\end{equation}

\noindent
with identical relationships for $G\left[D^{\bm{27}}_{(Y,I,I_3)}\to B'B\right] = G\left[D^{\bm{27}}_{(Y,I,I_3)}\to BB'\right]$.

\end{appendices}

\clearpage

\begin{table}
	
	\caption{
		Summary of studied decays and results found in this article. References to the corresponding sections and equations are shown.
		\label{tablai}}
	
	\begin{ruledtabular}
		
		\begin{tabular}{lcl}
			Decay & Independent amplitudes & Sum rules
			\\
			\hline
			$D_{\bm{10}}\to B_{\bm{8}}+R_{\bm{10}}$ & 13 & 8
			\\
			Sec.~\ref{sec.d10br} & Eqs.~(\ref{gd10br1}) - (\ref{gd10br7}) & Eqs.~(\ref{d10br1}) - (\ref{d10br8}) \\
			\hline
			$D_{\bm{27}}\to B_{\bm{8}}+R_{\bm{10}}$ & 22 & 16
			\\
			Sec.~\ref{sec.d27br} & Eqs.~(\ref{g27br1}) - (\ref{g27br11}) & Eqs.~(\ref{d27br1}) - (\ref{d27br16}) \\
			\hline
			$D_{\bm{27}}\to B_{\bm{8}}+B_{\bm{8}}$ & &
			\\
			Antisymmetric final state & 10 & 7
			\\
			Sec.~\ref{subsubsec.d27bba} & Eqs.~(\ref{g27bba1}) - (\ref{g27bba5}) & Eqs.~(\ref{d27bba1}) - (\ref{d27bba7}) \\
			\hline
			$D_{\bm{27}}\to B_{\bm{8}}+B_{\bm{8}}$ & &
			\\
			Symmetric final state & 14 & 10
			\\
			Sec.~\ref{subsubsec.d27bbs} & Eqs.~(\ref{g27bbs1}) - (\ref{g27bbs7}) & Eqs.~(\ref{d27bbs1}) - (\ref{d27bbs10}) \\
			
		\end{tabular}
		
	\end{ruledtabular}
	
\end{table}

\end{document}